\documentclass[review,3p]{elsarticle}
\pdfoutput=1

\usepackage{lineno,hyperref}
\usepackage[utf8]{inputenc} 
\usepackage[T1]{fontenc}    
\usepackage{url}            
\usepackage{booktabs}       
\usepackage{amsfonts}       
\usepackage{nicefrac}       
\usepackage{microtype}      
\usepackage{lipsum}
\usepackage{graphicx}
\usepackage[lofdepth,lotdepth]{subfig}
\usepackage{tabu}
\usepackage{subfig}
\usepackage{amsmath}
\modulolinenumbers[5]

\journal{Journal of Pattern Recognition}

\begin{document}

\begin{frontmatter}

\title{BERTERS: Multimodal Representation Learning for Expert Recommendation System with Transformer}

		
\author[mymainaddress]{N. Nikzad\textendash Khasmakhi }
\ead{n.nikzad@tabrizu.ac.ir}
\author[mymainaddress]{M. A. Balafar \corref{mycorrespondingauthor}}
\cortext[mycorrespondingauthor]{Corresponding author}
\ead{balafarila@tabrizu.ac.ir}
\author[mymainaddress]{M.\, Reza Feizi\textendash Derakhshi}
\ead{mfeizi@tabrizu.ac.ir}
\author[mysecondaddress]{Cina Motamed}
\ead{motamed@free.fr}


\address[mymainaddress]{Department of Computer Engineering, University of Tabriz, Tabriz, Iran}
\address[mysecondaddress]{Department of Computer Science, University of Orleans, Orléans, France}

\begin{abstract}
The objective of an expert recommendation system is to trace a set of candidates' expertise and preferences, recognize their expertise patterns, and identify experts.  In this paper, we introduce a multimodal classification approach for expert recommendation system (BERTERS). In our proposed system, the modalities are derived from text (articles published by candidates) and
graph (their co-author connections) information.
BERTERS converts text into a vector using the Bidirectional Encoder Representations from Transformer (BERT). Also, a graph Representation technique called ExEm is used to extract the features of candidates from co-author network. Final representation of a candidate is the concatenation of these vectors and other features. Eventually, a classifier is built on the concatenation of features. 
This multimodal approach can be used in both the academic community and the community question answering. To verify the effectiveness of BERTERS, we analyze its performance on multi-label classification  and visualization tasks. 
\end{abstract}


\end{frontmatter}


\section{Introduction}\label{sec:introduction}
Recently, the shadow of recommendation system (RS) has appeared on various domains and applications. On the other hand, significant new advances in deep learning approaches have important effects on the tremendous success of the recommendation system \cite{Zhang2019}. The overall structure of a RS follows a set of phases including collection, learning, and recommendation \cite{Bobadilla2013, Isinkaye2015}. In the first phase,  appropriate
resources that comprise the relevant information of users are selected. Then, a leaner (supervised or unsupervised learning) analyzes the users’ preferences, and extracts their behavioral patterns. Final phase recommends the entities that are the most similar to the users' interests. It is important to recognize that, within a common core structure of RS, there are variations from
application to application. Some of the most sophisticated and heavily used RSs in industry are Last.fm, YouTube, and Amazon.

Furthermore, we can find the footprint of RS in the knowledge management system where RS tries to specify experts who have the most relevant knowledge about a particular topic \cite{Zhen2012, nikzad2019state}. This category of RS is called expert recommendation system (ERS) or expert finding system. So, it is obvious that an ERS has similar phases compared
to general RSs. An ERS takes a user topic or query, traces a set of candidates' expertise, learns their expertise patterns, and finally produces a list of experts sorted by a score. Each candidate's score indicates the degree of this candidate's relevant expertise with the given topic. 
In the most studies, the candidates' expertise is defined as content-based information and non-content-based information \cite{Wang2013}. Content-based information is
candidates' shared textual content like their articles, questions, answers and so on. In contrast,
candidates' interactions with each others in social networks make non-content-based information. Depending on the application scenarios, each ERS has its own set of contextual information. For example, in academic environment, the attempt is to detect researchers who have the subject areas related to the query. This detection is based on the content of the articles published by them and their co-author relations in different papers. However, in Community Question Answering (CQA), the main goal is to find the users with expertise and willingness to answer the given questions in terms of the content of the question asked and the answered posted by them, and their question-answer relations \cite{DBLP:journals/corr/abs-1804-07958}.

With a brief look at previous studies, it can be concluded that there are three different outlooks on ERSs. In one of the attitudes, studies have focused on the textual expertise of candidates. These works have used text mining or information retrieval techniques and selected ones as experts who their published items are semantically relevant to the query \cite{Mumtaz2019}. On the other hand, some other researches have investigated the social relations between candidates and represented their connections as a graph \cite{Yang2008}. After that,  social network analysis and mining techniques, such as page ranking algorithms, are applied on this graph to identify important candidates and rank them. Moreover, recent studies have shown that the combination of different types of expertise information have notable performance compared to other. A number of them have integrated textual expertise and social network connection information with linear or nonlinear functions. 
Also, to achieve higher accuracy, a few authors proposed the usage of heterogeneous network which is a combination of the users' interactions in social networks and their question–answer relationships in CQAs besides bearing in mind the content of questions and answers.

In recent years, multimodal machine learning has been attracting attention. This popularity is because of huge multimodal content being generated by the users of social media networks \cite{Baltrusaitis2019} . The goal of multimodal machine learning is to create a joint model that can retrieve contextual information from multiple modalities \cite{Atefeh2015}. In this research, we aim to find academic experts that whether using a multimodal learning approach provides an effective solution for ERS or not. Also, the other purpose of our work is to solve the expert finding problem as a multi-label classification task. In such a way, we combine text (articles) and
graph (co-author connections) information in a multimodal approach. The text component is converted into vector using BERT Transformer. On the other hand, to obtain the node representation of candidates, a graph embedding technique, ExEm introduced in \cite{nikzad2020exem}, is used. Also, normalized h-index value of candidate is added as another feature. Then, the captured fusion features are fed to train the classifier. We evaluate BERTERS on the multi-label classification  and visualization tasks. However, to the best of our knowledge, we present the first approach in field of expert recommendation using multimodal learning and transformers.

The rest of the paper is structured as follows: Section \ref{related_work} reviews the related works. Section \ref{backgorund} discusses the background of the research. Section \ref{Proposed_Method} presents our proposed method and explains it in detail. 
 The descriptions of the dataset and the tasks that are used to test our proposed method and parameter setting are presented in Section \ref{experiments}.  Section \ref{evaluation_re} provides and discusses the experimental results. Finally, Section \ref{sec:conclusion} concludes the paper.

\section{Related Work}\label{related_work}
In this section, we review the approaches proposed for ERS. We group these models into the three categories, based on their main outlooks: document-based, graph-based and hybrid models. The bellow subsections will explain the underlying method logy and existing approaches for the specified categories. In case of reader curiosity, we highly recommend reading \cite{nikzad2019state} that explains in more detail and is dedicated to review all the related articles in this scope.

\textbf{\subsection{Document-based models}} 
Document-based models are intended to compare the characteristics of the content contained in the published items associated with a candidate and the query. On the other hand, document-based models work well where capturing the level of expert’s knowledge in the field of the topic query is the goal. 
A number of works employed topic modeling techniques for this task. In study \cite{Riahi2012}, authors suggested a framework to automatically direct new questions to the best experts based on tracking their answering history in the community. Their proposed solution employed different methods consisting language models with Dirichlet smoothing, TF-IDF, Latent Dirichlet Allocation (LDA) and Segmented Topic Model for this aim. Research \cite{Momtazi2013} applied LDA method to collect the topics of  documents. After that, the probability of each candidate query is calculated based on the extracted topics for each query. Experts are sorted according to this probability. In another paper, Neshati et al. \cite{Neshati2017dy} emphasized on the dynamic aspects of the expert finding.  Authors considered four content features including topic similarity, emerging topics, user behavior and topic transition features to predict the best ranking of experts in future. 
There are other interesting document-based models. Nobari et al. \cite{Nobari2017}  proposed two translation models  based on a statistical approach and a word embedding model. \cite{Li2015} presented a tag-LDA approach to model the candidate topic distribution. Despite the fact that document-based approaches are helpful in finding the knowledgeable candidates, they can not detect the important or influential experts in the social network.

\textbf{\subsection{Graph-based models}} 
Document-based models recognize expertise patterns across documents, whereas graph-based approaches learn to recognize patterns across graph. Graph-based models work well where  authority and reputation scores of candidates are important. Authority score measures the influence
and popularity of candidates in social networks. On the other hand, candidates with high reputation scores share more knowledge and information with others in the communities \cite{nikzad2019state}.
The graph-based method formulates the problem of ERS from the perspective of a graph $G(V, E)$, where $V$ denotes a set of candidates and $E$ a set of edges among the nodes. Depending on the applications at hand, nodes can represent candidate experts of various types such as academic candidates. On the other hand, edges represent different types of relations between any candidates,
such as question posters and repliers relations in CQA or follower-following connections in social networks, etc.  Most previous graph-based methods were used PageRank and HITS, two popular link analysis approaches to measure the similarity between candidates with a topic query, calculate candidates' scores and  make recommendation. Fu et al. \cite{Fu2007} proposed an expertise propagation algorithm that is very similar to PageRank to build the relationship between candidates. Zhang et al. \cite{Zhang2007} used the authority value of HITS algorithm to select a user as expert who helps many others. Also, authors introduced ExpertiseRank, an algorithm similar to PageRank, to measure experts' authorities \cite{Lin2017}. Also, there are some other papers focusing on detecting the top-K influential candidates in communities \cite{Zhan2016}. Mumtaz and Wang \cite{mumtaz2017identifying} proposed a simple technique to find the influential node set in a network with largest betweenness centrality. Paper \cite{Bian2019} reviewed the existing works on identifying top-k influential and significant nodes. Although, the graph based approaches find the influential candidates in the social network, they fail to consider each expert candidate's topical expertise.

\begin{figure}[ht]
    \centering
    \includegraphics[page=1,width=0.8\linewidth]{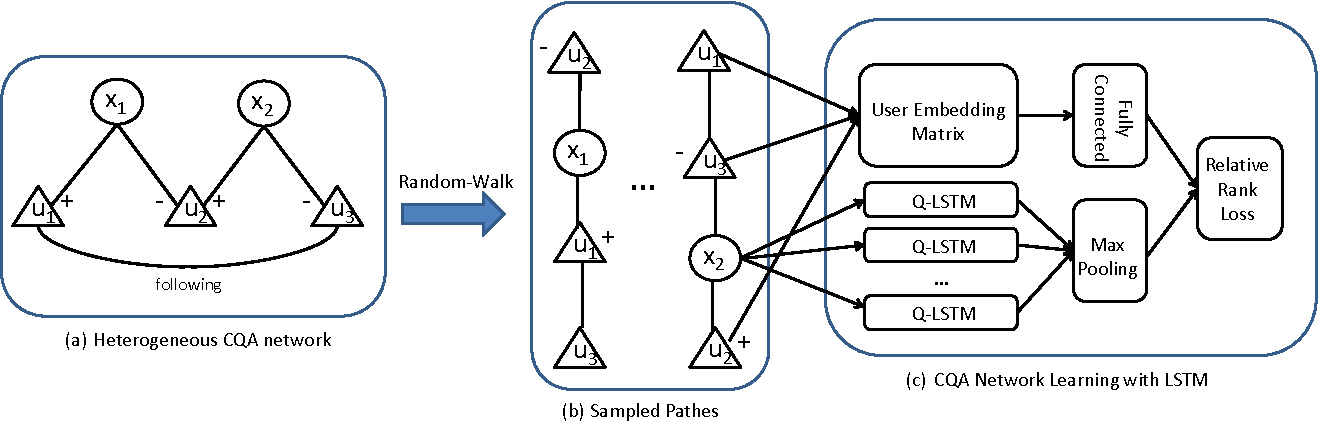}
    \caption{The architecture of ranking metric network learning framework \cite{Zhou2016}.}
    \label{fig:RMNL}
\end{figure}

\textbf{\subsection{Hybrid models}} 
Hybrid models have drawn a lot of attention for ERS in recent years. These methods have been developed to combine features extracted from the documents (or questions and answers), and features obtained from  candidates' social network communications to formulate a recommendation. It should be noted that hybrid models need to use a feature-combination method to merge content and non-content expertise and calculate scores. This section reviews some of the most prominent hybrid models which created new state-of-the-arts on ERS. Zhao et al. \cite{Zhao2015} proposed a hybrid model (GRMC) created from both the social relationship between candidates and their history of questions and answers. In proposed model, the goal is to consider expert finding as missing value estimation and estimate values via a matrix completion method. In \cite{Zhou2016}, Zhao et al. proposed a ranking metric network learning framework (RMNL) for the problem of expert finding. As illustrated in Fig \ref{fig:RMNL}, they performed a heterogeneous CQA network built by the combination of both candidates' relative quality rank to questions and their social connections. Sang et al. \cite{Sang2019} proposed a hybrid model (MMSE) which is similar to GRMC in \cite{Zhou2016}. Authors designed a bayesian embedding model which integrates multiple modalities and multiple semantic perspectives. Zhou et al. \cite{Zhou2014} considered the candidate expertise and reputation score for finding experts. They proposed a user-topic model to analyze the content of
the questions and answers. Moreover, authors introduced a topic-sensitive method to reflect both the link structure and the topic relevance between questioners and answerers. In \cite{Liu2013}, Liu et al. merged knowledge, reputation and authority scores of candidates to produce a recommended expert list. Knowledge score shows the similarity of the profile and the target question. Moreover, the number of answers and best answers given by candidates  are used to find the reputation score. Finally, the authority score is calculated using HITS and Page Rank approaches. Xie et al. \cite{xie2016topic} used LDA and HITS algorithms
to extract topical feature. The suggested method evaluated social relation, time and location
factors in order to extract contextual features. Finally, a SVM algorithm was used as scoring function.  There are other interesting hybrid models including CQARank \cite{Yang2013}, ExpertRank \cite{Wang2013}, Expert2Vec \cite{Mumtaz2019}. The  hybrid models have achieved high accuracy on many ERS benchmarks. But, the important point in these approaches is how to combine text and link elements to detecting experts.

\section{Background}\label{backgorund}
In this section, we discuss the concepts which organize the background of our
study. In this way, firstly, the text representation method, BERT Transformer, is explained. After that, the graph embedding technique, ExEm, is introduced.

\subsection{Text representation learning}\label{sec:bert}
\begin{figure}[ht]
    \centering
    \includegraphics[page=1,width=0.5\linewidth]{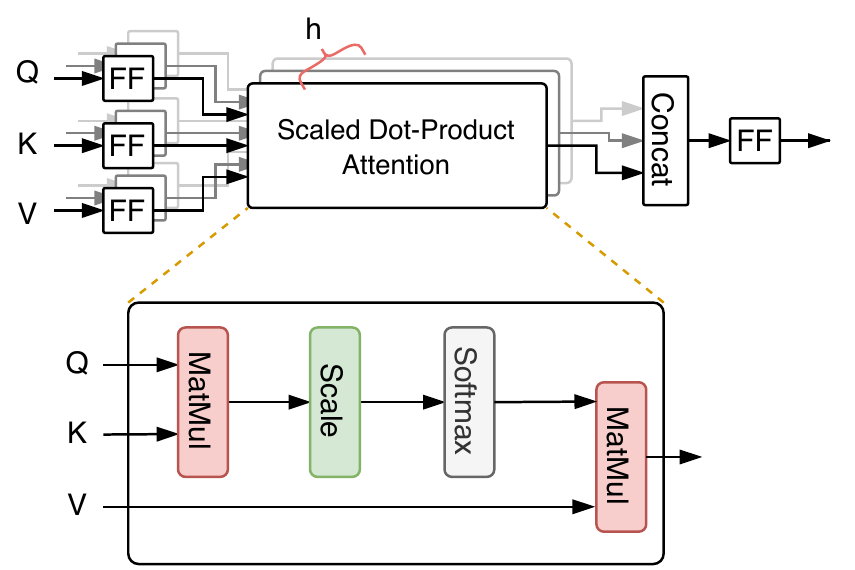}
    \caption{ The Transformer model architecture \cite{asgari2020multimodal}}
    \label{fig:transformer}
\end{figure}

In recent years, researchers have devoted many efforts on extracting features from text data, and have proposed many models including neural embedding, attention mechanism, self attention, and Transformer. As investigated in many papers, the sequential processing of text and the computational cost of obtaining remarkable relationships between words in a
sentence are two issues that RNN and CNN models are encountered with, respectively. On the other hand, Transformers eliminate these bottlenecks by assigning in parallel an attention score to each word in a document to consider the impact of words on each others \cite{minaee2020deep}. Fig \ref{fig:transformer} illustrates the architecture of  the Transformer model that comprises of both  encoding and decoding components which are all identical in structure. These components include the stacked layers. For example, the encoding component is a stack of encoders where each stack layer is broken down into two sub-layers. Each sub-layer has a multi-head attention layer and a feed-forward neural network. The multi-head attention layer extracts the dependencies between representation pairs regardless of the distance between them in the sequence and is more effective than single-head attention \cite{minaee2020deep,Sun2019}. The outputs of the the attention layer are injected to the feed-forward. For each set of queries $Q$, keys $K$ and values $V$, the multi-head attention module applies $h$ attention functions which is the scaled dot-product attention as shown in equation \ref{eq:transformer}.

\begin{equation}\label{eq:transformer}
Attention(Q, K, V ) = softmax(\frac{QK^T}{\sqrt{d_k}})V
\end{equation}

\begin{figure}[ht]
    \centering
    \includegraphics[page=1,width=0.7\linewidth]{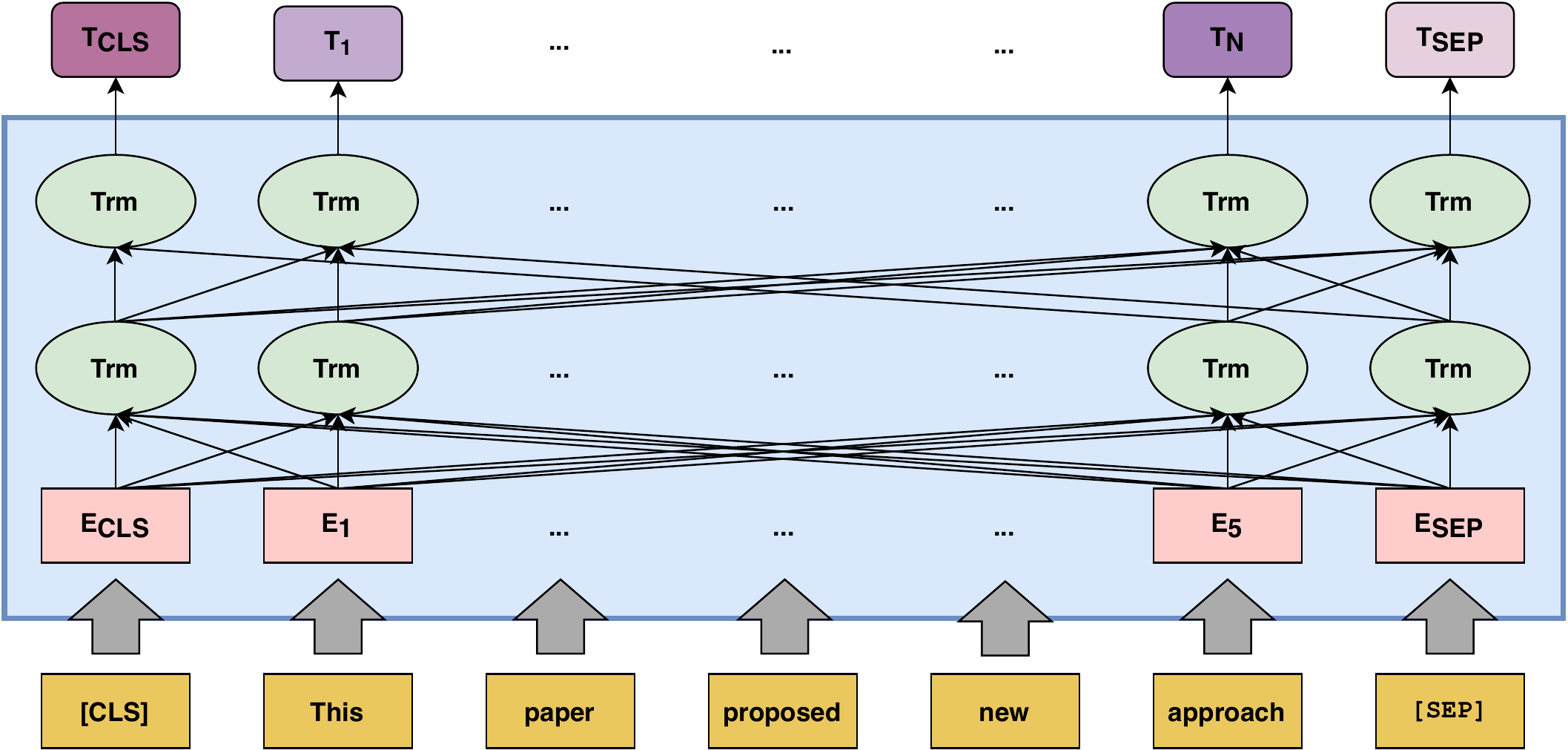}
    \caption{The BERT Transformer architecture.}
    \label{fig:bert}
\end{figure}

One the most widely used Transformer models is BERT Transformer \cite{devlin2018bert} that is the new state-of-the-art sentence embedding models \cite{minaee2020deep}. The BERT Transformer architecture is shown in \ref{fig:bert}. A masked language modeling task is used for training BERT. It randomly selects some tokens
in a text sequence for masking, and then independently retrieves the masked tokens by conditioning on the encoding vectors which are the outputs of a bidirectional Transformer. For using BERT, firstly, two tokens, that are known as $[CLS]$ and $[SEP]$, are added at the beginning and the end of the text input, respectively. After that, the input flows through the two transformer layers. The output of the last transformer layer is the embedding of the input. Briefly, BERT model has two parameters $h$ and $L$. $h$ is the size of the output embedding vector and $L$ shows the number of stacked layers in each component.

\subsection{Node representation learning}\label{sec:exem}
One of the key concepts in the analysis of social networks is the idea of presenting the knowledge inside them as a graph structure \cite{Nettleton2013}. On the other hand, in the recent times, one of the most widely used graph analysis approaches is graph embedding.  Graph embedding represents the graph nodes as low-dimensional vectors \cite{Cai2018,Goyal2018}.  It gives us a deeper vision to analyze users' activity patterns and their relationships in social networks.  A number of recent techniques have developed  to embed graph nodes.  In our study, we focus on three embedding techniques including DeepWalk \cite{Perozzi2014}, Node2vec \cite{Grover2016} and ExEm \cite{nikzad2020exem} that employ random walks on a graph to obtain node representations.  

DeepWalk is the first effort proposing the deep learning techniques into graph analysis. Because the random walks can govern the structure of graph, DeepWalk uses a stream of short random walks. It considers each random walk as a sentence, and the graph nodes as words. Therefore, authors generalized the idea of language modeling in NLP to explore the graph. The aim of language modeling is compute the probability of a sentence
or the sequences of words as shown in equation \ref{language_model}.

\begin{eqnarray}\label{language_model}
P(w)=P(w_1w_2...w_m)=\prod_{i=1}^mP(w_i|w_1w_2...w_(i-1)) = P(w_1)P(w_2|w_1)P(w_3|w_1w_2)...P(w_m|w_1...w_{m-1})
\end{eqnarray}

To transfer the language modeling into the graph, the task is to estimate the probability of equation \ref{language_model_node}.

\begin{eqnarray}\label{language_model_node}
P(v_i|(v_1v_2...v_{i-1})) = P(v_i|\Phi(v_1)\Phi(v_2)...\Phi(v_{i-1})) = \prod_{i=1}^mP(v_i|\Phi(v_1)\Phi(v_2)...\Phi(v_{i-1}))
\end{eqnarray}
where  $\Phi$ is the low-dimensional representation of each node in the graph.

In Node2vec, authors introduced a flexible strategy to generate the node's neighborhood. They designed a biased random walk procedure based on the concept of the breadth-first and depth-first search algorithms. In order to bias the random walks, two parameters $p$ and $q$ control the likelihood of immediately revisiting a node in the walk and the distances from a given source node, respectively. Node2vec uses an extended version of the Skip-gram architecture to optimize the stochastic gradient descent.

ExEm is another graph embedding technique that applies the dominating set theory on the graph and finds the dominating nodes. Then, ExEm creates a set of random walks that contains at least two dominating nodes, and stores it as a corpus. In the next step, the corpus is fed to Word2vec, fastText and their combination  to train the Skip-gram neural network.

\section{Proposed Method}\label{Proposed_Method}
\begin{figure}[ht]
    \centering
    \includegraphics[page=1,width=0.4\linewidth]{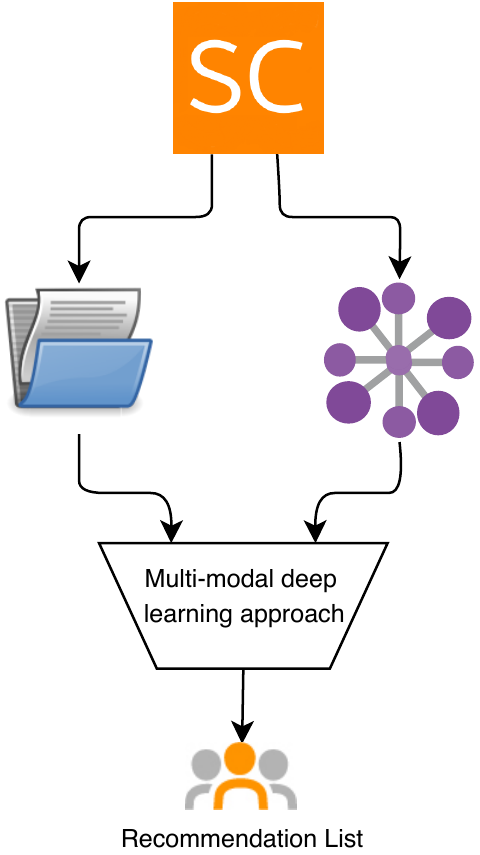}
    \caption{The overall structure of BERTERS.}
    \label{fig:BERTERS}
\end{figure}

The aim of this paper is to design a new hybrid model with a multimodal neural network, which is called BERTERS, that is able to find academic experts. The overall structure of BERTERS is shown in Fig \ref{fig:BERTERS}. In the first step, we extract the adequate dataset from \href{https://www.scopus.com/}{Scopus} which is the largest abstract and citation database. The gathered dataset includes the content and non-content features of expert candidates such as their published articles, subject areas, affiliations, h-index, and their co-author interactions. In the next phase, BERTERS takes as input the articles and the co-author connections that have various types (e.g., text and graph). Hence, these different modalities enable a multimodal deep learning approach to create comprehensive and meaningful representations of expert candidates.  To capture candidates' representations from these different modalities, BERTERS is comprised of three different neural networks: one for document representation generation, the other one for node representation generation and the third one for learning a shared representation between modalities. Each feature is separately obtained from the respective neural network and then merged with other features to create a single representation for each candidate. Finally, the model provides a list of candidates as experts via collaborative filtering.

To the best of our knowledge, BERTERS is the first recommendation model for ERS that employs multimodal learning and transformers. Although MMSE \cite{Sang2019} proposed a multimodal approach for finding experts in CQA, but BERTERS is the first use of the multimodal classification approach in the context of ERS in an academic community. As of another meaning, BERTERS perceives the ERS as a vision of a multi-label classification task using multimodal learning. In MMSE, authors used the Skip-gram model and DeepWalk to learn word embeddings and network-based user embeddings, respectively.  Conversely,  BERTERS employs BERT transformer and ExEm method to capture document and graph embeddings, accordingly. Also, our approach adds candidate's h-index as another feature.  The following subsections describe the procedures of BERTERS in detail.  

\subsection{Model Architecture}
\begin{figure}[ht]
    \centering
    \includegraphics[page=1,width=1\linewidth]{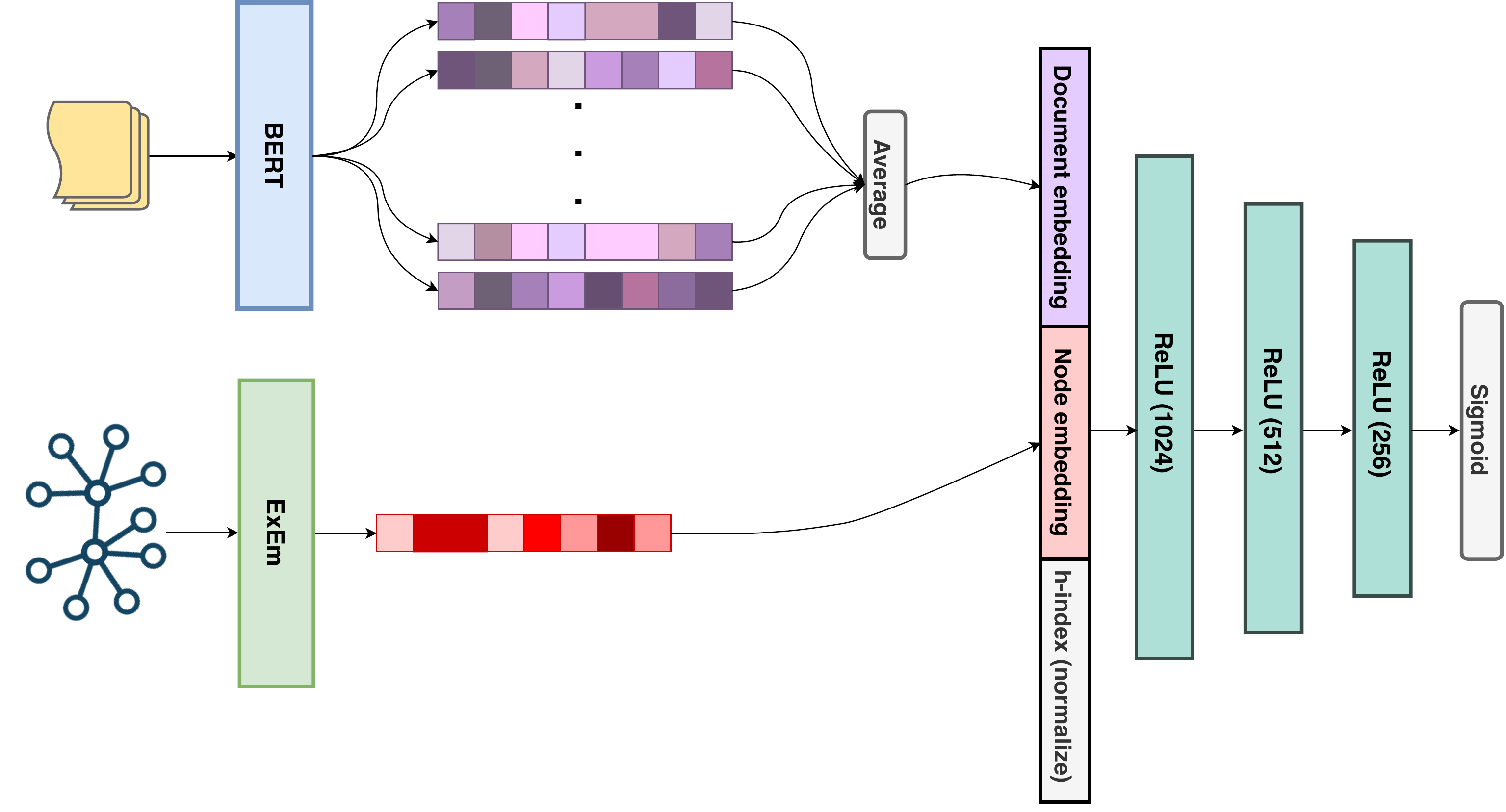}
    \caption{multimodal architecture of BERTERS}
    \label{fig:multimodal}
\end{figure}

As it was mentioned previously, in this study, we introduce a multimodal deep learning approach that considers the ERS as a multi-label classification task, shown in Fig \ref{fig:multimodal}. From this viewpoint, the prediction problem becomes accurately classifying a specific expert candidate where the candidates' subject areas are defined as their labels. This model can be formalized as computing
the probability all possible subject areas for an expert candidate based on the average of all document embedddings $D_e$, candidate social connection embeddding $N_e$, and h-index $Hi$:

\begin{eqnarray}\label{eq:classify}
P(C_{sa}|D_e, N_e, Hi) = P(v_i|[D_e; N_e; Hi]) 
\approx P(C_{sa}|E)
\end{eqnarray}

where $E$ is defined as the concatenation of $D_e, N_e, Hi$ and applying three dense layers with ReLU function. In other sense,  the task is to learn expert candidate embeddings $E$ as a function of articles, co-author relations and h-index that is presented in equation \ref{eq:classify_2}. 

\begin{eqnarray}\label{eq:classify_2}
E = ReLU(ReLU(ReLU([D_e; N_e; Hi]W)W)W)
\end{eqnarray}

The direct analog is to estimate the likelihood of subject areas of a candidate based on $E$. Hence, a Sigmoid classifier applies on the embedding $E$. Equation \ref{eq:classify_3} shows this probability.

\begin{eqnarray}\label{eq:classify_3}
P(C_{sa}|E) = Sigmoid(E)
\end{eqnarray}

\subsection{Document representation generation}
As it can be concluded from Fig \ref{fig:BERTERS}, one of the BERTERS modalities is text information that comes from the articles published by candidates. It aims at extracting distinguishing text expertise of candidates. As presented in Fig \ref{fig:multimodal}, we learn the representation of each document in a fixed-sized via BERT Transformer demonstrated in section \ref{sec:bert}. The input of BERT is an article of each candidate. The article passes through the layers of BERT, and the output is its embedding. Consequently, a candidate's content information is represented by a high-dimensional vector, $D_e$, which is the average of his/her all article embeddings. 

It is worth noting that we can extend this procedure for the ERS in CQA. For this aim, the questions asked and the answers posted by candidates are fed into inputs of the BERT model. After that, the average of these embeddings are used as the text modality value.

\subsection{Node representation generation}
Learning features of modalities is the foundation of multimodal deep learning approaches. As explained before, another modality in BERTERS fetches from candidates' co-author network. To interpret information of this network, we use the graph embedding techniques, DeepWalk, node2vec, and ExEm that are described in section \ref{sec:exem}. The candidate' node embedding representation $N_e$ is generated by applying the graph embedding method on the collaborative network.

In order to apply this strategy in CQA, the desired graph is constructed based on the interactions between question posters and repliers. Other steps are done as described above.

\subsection{Other features}
Adding features results in having a depth knowledge about candidates' expertise, accurately learning their subject areas, and improving precision. Hence, we also add h-index of candidates in form of additional feature. On the other hand, the proper normalization of features is critical for convergence. So, h-index is normalized, defined as $Hi$, to combine with the features obtained from previous stages.

To use BERTERS in CQA system, we can add number of best answers provided by candidates, their reputation score, number of thumbs up and down as extra features.

\subsection{Joint Features}
The important point in a multimodal deep learning model is to properly integrating multimodal features. But in practice, combining different modalities is challenging. Furthermore, modalities have different quantitative effects on the prediction output. There are at least three common ways to combine embedding vectors a single feature vector including: summing, averaging and concatenating \cite{Damoulas2009}. 

In our case study, because the length of modality representations are not the same, it is not possible using summing and averaging methods. In this way, we integrate all features into a single representation through concatenation and get $1 \times L$ vector, where $L$ equals to the sum of the length of feature vectors. In the next step, BERTERS employs a feed-forward neural network which consists of three stacked dense layers with Rectified Linear Units (ReLU) activation function. The last layer is Sigmoid classifier.  To efficiently train BERTERS, a cross-entropy loss is minimized and embeddings are learned jointly with all other model parameters.


\section{Experiments}\label{experiments}
In this section, we present the details of the experiment process. We start with explaining the dataset and how it is obtained and the related information, later we jump into the experimental setup of our work and then tasks, model variation and metrics are described in order.

\subsection{Dataset}

To evaluate the performance of BERTERS, we search for a dataset which guarantees both content and non-content modalities. The dataset introduced in \cite{nikzad2020exem}, gathered from Scopus eliminates the require to a labeled data for constructing a collaborative network. The graph extracted from this dataset has arisen out of the collaborations of authors in different articles. Each node presents an author that his/her subject areas are considered as node labels. Moreover, the edges indicate the co-author interactions between authors.  This dataset only ensures the data of graph modality. To adapt this dataset to our multimodal approach, we extract some other features from Scopus for text modality. The obtained information consists of authors' articles, their h-index and affiliations.
An important point about the dataset which our experiments use, is that the total number of the graph nodes is 27,473, but the text information is gathered only for 9,378 authors.
The descriptions of the dataset is summarized in Table \ref{tb:dataset}.


\begin{table}[h]
\centering
\caption{Dataset information.}
\label{tb:dataset}
\begin{tabular}{c|c|c|c|c}
\hline
$|V|$ & $|E|$ &  Labels & $\#$ Articles &  $\#$ Authors with articles  \\
\hline
27,473 & 285,231 & 27 & 472,566  & 9,378 \\
\hline
\end{tabular}
\end{table}

\begin{figure}
\centering
    \subfloat[IE topic]{{\includegraphics[width=0.32\linewidth]{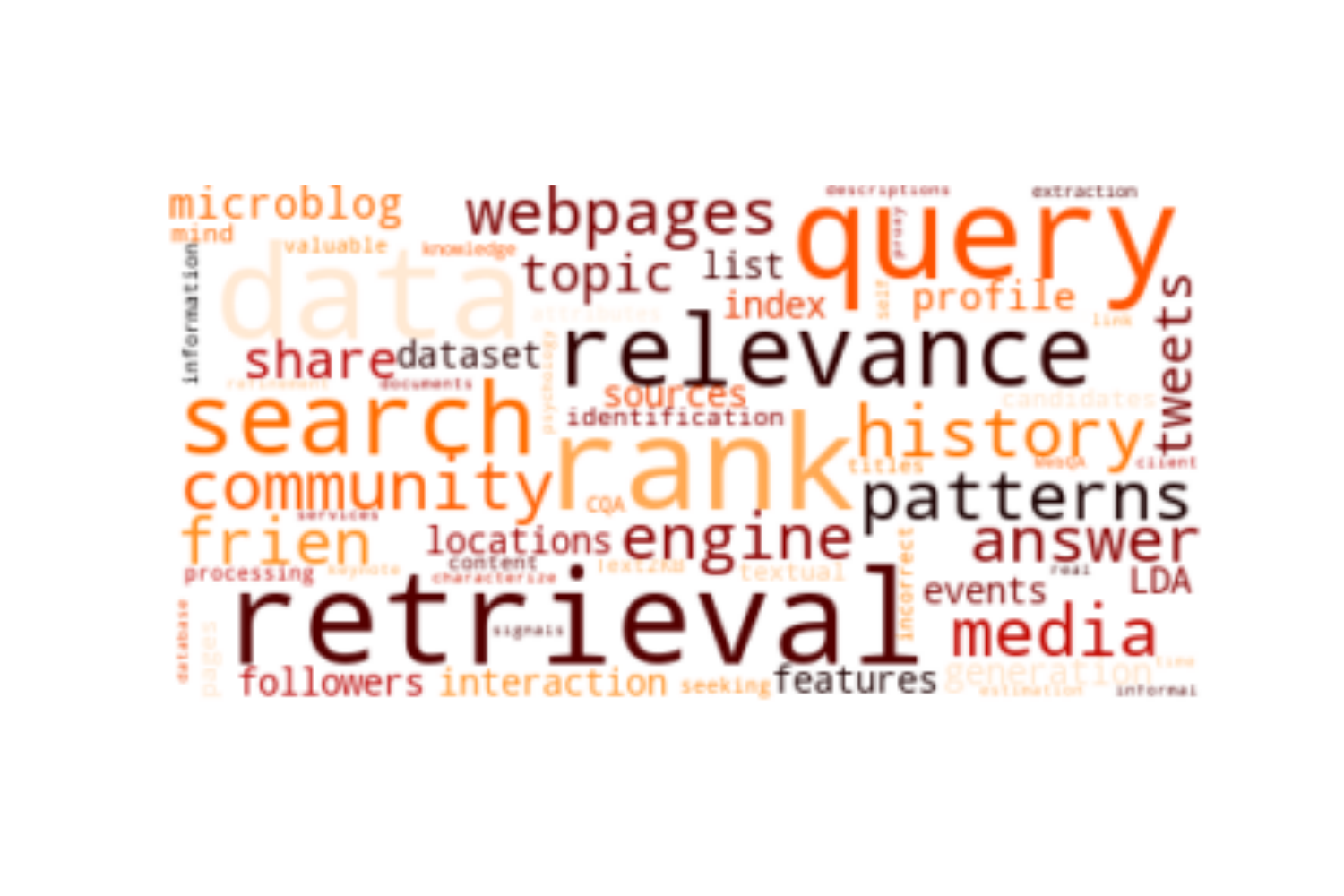} }}%
    \hfill
    \subfloat[ML topic]{{\includegraphics[width=0.32\linewidth]{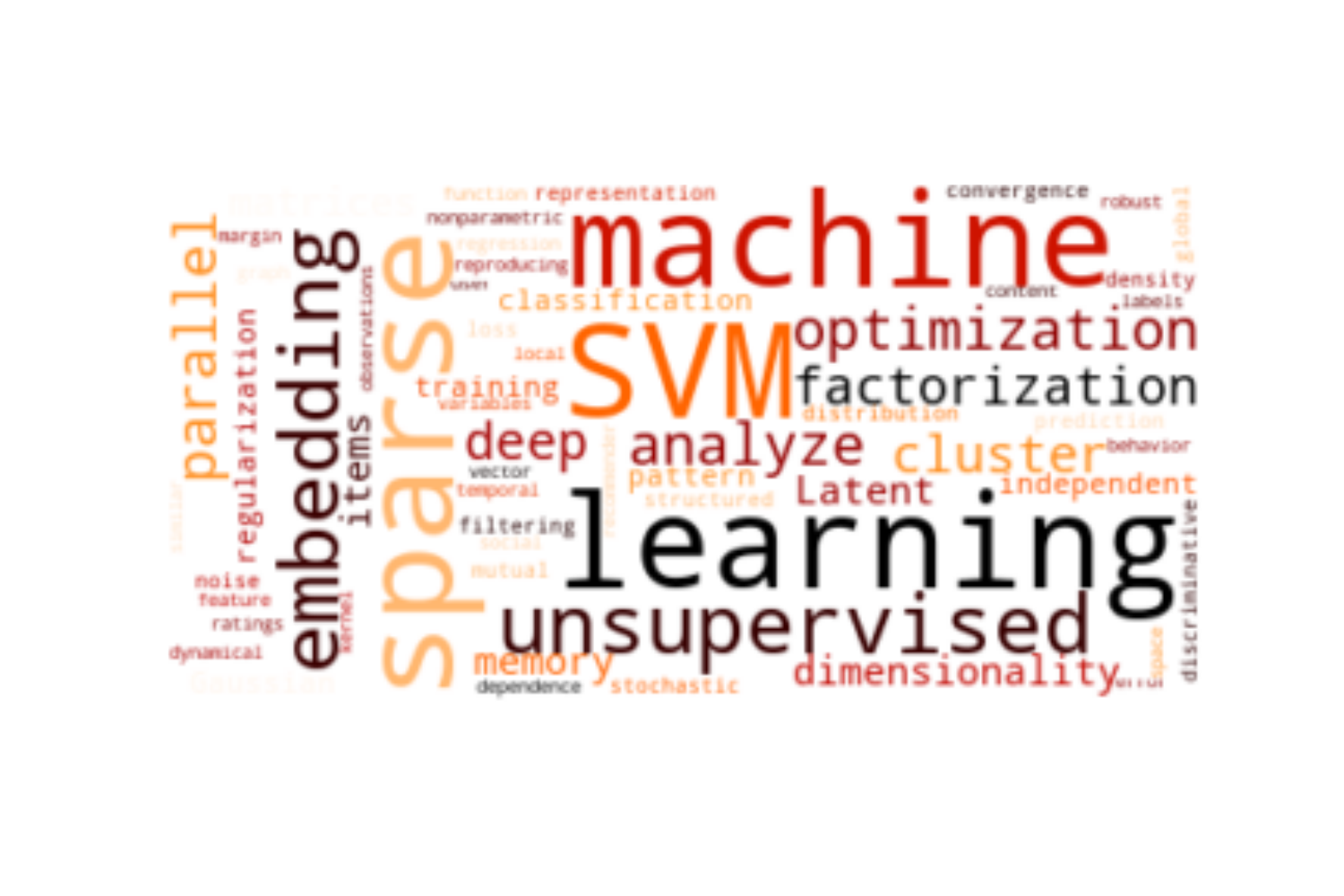} }}%
    \hfill
    \subfloat[NLP topic]{{\includegraphics[width=0.32\linewidth]{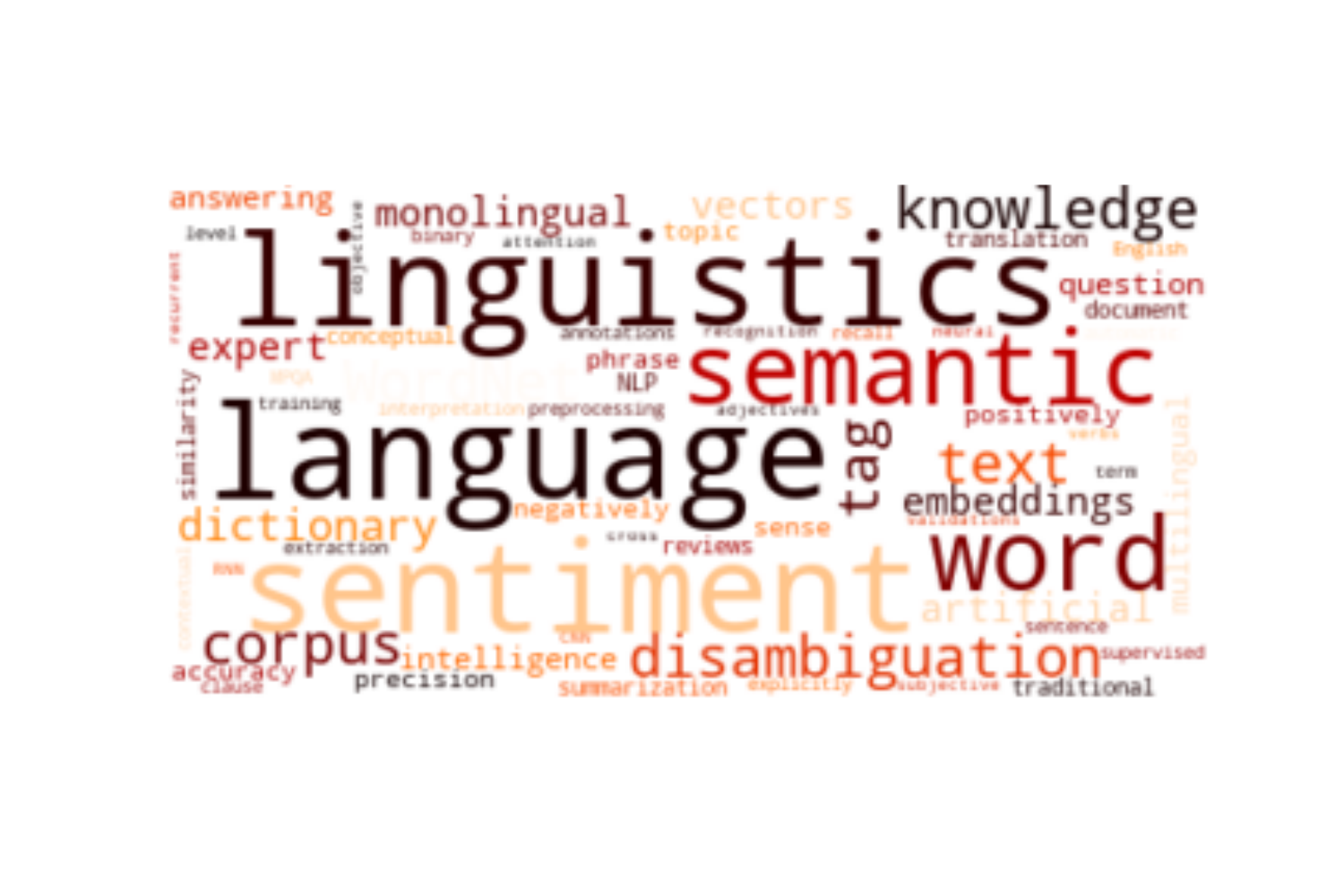} }}%
    
\caption{Word cloud presentation of articles related to the top experts for three topics}\label{fig:dataset}
\end{figure}

Because BERTERS is a supervised multimodal classification approach, so it needs a ground truth for learning. Hence, to find a proper ground truth for our collected dataset, we follow the same procedure described in \cite{nikzad2020exem}.  We derive a list of experts  from \href{https://aminer.org/lab-datasets/expertfinding/}{\textit{Arnetminer}} for three topics: information extraction (IE), natural language processing (NLP), and machine learning (ML).  This list of experts and the topics are defined as ground truth and query, respectively. Fig \ref{fig:dataset} shows the word cloud presentation of the articles related to the top expert in three topics.

\subsection{Experimental Setup}
In our study, we employ a version of BERT called BERT-Small. Its encoding and decoding parts have 4 stacked layers. Also, the size of the output embedding vector in BERT-Small is 512.  The required information about setup is denoted in Table \ref{tab:ex_setup}. Furthermore, Table \ref{tb:system_info} presents the information of system that the experiments were performed on.

\begin{table}[ht]
    \centering
    \begin{tabular}{l|c}
        BERT embedding vector size  & 512 \\
        ExEm embedding vector size &  128 \\
        h-index feature size & 1 \\
        Total embedding size & 641 \\
        Number of classes & 27 \\
        Number of clusters & 3 \\ 
    \end{tabular}
    \caption{Experimental setup}
    \label{tab:ex_setup}
\end{table}

\begin{table}[ht]
\centering
\caption{System Information.}
\label{tb:system_info}
\begin{tabular}{l|c|c}
\hline
 &  Model  &  Description  \\
\hline
OS& Ubuntu 18.04.3 LTS  & -	\\  \hline 
RAM & - &  26G  \\ \hline
CPU & Intel(R) Xeon(R)  &  2.20GHz  \\ \hline
GPU & NVIDIA  &  Tesla P100 \\ \hline
\end{tabular}
\end{table}

\subsection{Tasks}
We evaluate the performance of BERTERS on two tasks including multi-label classification  and visualization that are described in the following.
 
\subsubsection{Multi-Label Classification}
In this task, the effort is to predict the labels of candidates with high precision. In our work, the labels of candidates are defined according to their subject areas and represented as a one-hot numeric array. 

\subsubsection{Visualization} 
Visualization assists in the achievement of more vision into the structure of the network. BERTERS illustrates the goodness of its embedding approach by clustering together experts based on three topics.

\subsection{Model Variations}
We experiment with several variants of the model.

\textbf{BERT}:  This model only operates on authors' articles. Each article is presented by a vector created form BERT transfromer.

\textbf{ExEm(fastText)}: It is a version of ExEm that engages fastText method to learn the node representation.

\textbf{ExEm(Word2Vec)}: This one is another form of ExEm that allows to create vector representations for nodes by using Word2Vec.

\textbf{BERTERS(ExEm(fastText))}: It is the combination of text and graph modalities. Text features are obtained by BERT transformer from articles. On the other side, ExEm(fastText) extracts node features from co-author graph.

\textbf{BERTERS(ExEm(Word2Vec))}: Same as above but the node vectors are captured by ExEm(Word2Vec).

\textbf{BERTERS(Node2Vec)}: This architecture is almost identical to the previous one. The difference is that Node2Vec approach creates the node vectors.

\textbf{BERTERS(DeepWalk)}: In this structure, DeepWalk derives the nodes features. The rest of the procedure is similar to the other BERTERS variations.

\subsection{Evaluation Metrics}\label{sec:eval}
The main metric which is used to evaluate the micro and macro F1 score that is expressed by the equation \ref{eq:f1}. From this equation, $Pr$ denotes precision and $Re$ denotes recall. However, this form of $F1$ is for general propose and not for macro and micro. The macro and micro $F1$ is expressed by using micro and macro precision and recall instead. 

\begin{equation}\label{eq:f1}
    F1 = 2\times\frac{Pr\times Re}{Pr + Re}
\end{equation}

Using micro and macro evaluation metrics makes sense in using multilabel or multi-dataset evaluations. In our work, a multilabel task has been proposed and accordingly, the micro and macro F1 should be reported. For computing the precision and recall, in micro, equations \ref{eq:micropr} and \ref{eq:microre} express the mathematical definition; For the macro precision and recall, \ref{eq:macropr} and \ref{eq:marcore} present definitions respectively.

\begin{equation}\label{eq:micropr}
    microPr = \frac{\sum\limits_{c\in Classes} TP_c}{\sum\limits_{c\in Classes}{(TP_c+FP_c)}}
\end{equation}

\begin{equation}\label{eq:microre}
    microRe = \frac{\sum\limits_{c\in Classes} TP_c}{\sum\limits_{c\in Classes}{(TP_c+FN_c)}}
\end{equation}

\begin{equation}\label{eq:macropr}
    macroPr = \frac{1}{N(Classes)} \sum\limits_{c\in Classes} Pr_c
\end{equation}

\begin{equation}\label{eq:marcore}
    macroRe = \frac{1}{N(Classes)} \sum\limits_{c\in Classes} Re_c
\end{equation}

\section{Results}\label{evaluation_re}
In this section, we investigate the efficiency of different embeddings on the tasks presented above. We also present the effect of number of embedding dimensions on the performance for each task.

\subsection{Multi-Label classification}
Evaluation and comparison of our proposed models is acquired by using the equations from subsection \ref{sec:eval}. Macro and micro F1 score of our method with its different variations are presented in tables \ref{tb:microf1} and \ref{tb:macrof1}. Utilization of text modality with the representation obtained from graph by different graph representation learners and our trained BERT model, shows that our hypothesis about using multimodal learning and obtaining better results is true. However, based on the learner itself that is the base for graph representation learning, BERTERS(ExEm(fastText)) is far better than others.

As it can be concluded from the tables, two single-modality based methods, BERT and ExEm produce poor consequences. However, employing document embeddings built by BERT presents better outcomes than node embeddings obtained from ExEm. In contrast,
using the multimodal approach can significantly improve the performance of ERS than single modal. Among variants of BERTERS, BERTERS(ExEm) achieves high micro and macro values in most cases. It comes from the fact that ExEm effectively monitors the network by help of dominating nodes. Although, DeepWalk and Node2vec also are random walk based methods but their walks do not provide enough information about nodes \cite{nikzad2020exem}. On the other hand, 
the results prove the efficiency of BERTERS(Node2Vec) than BERTERS(DeepWalk) due to designing biased random walk procedure.

\begin{figure}[ht]
    \centering
    \includegraphics[page=1,width=0.5\linewidth]{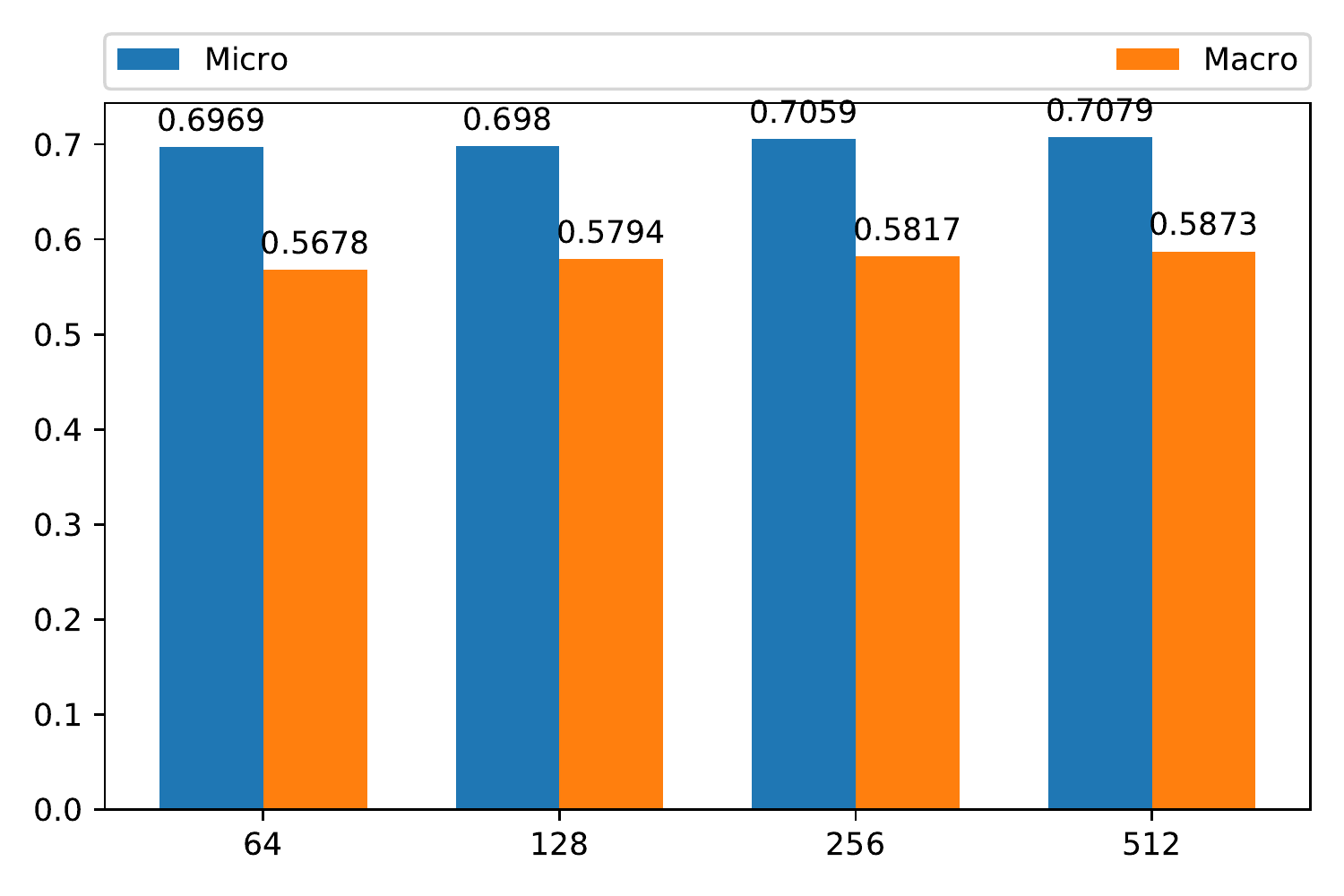}
    \caption{Micro-F1 and Macro-F1 of multi-label classification task for BERTERS(ExEm(fastText)) varying the number of dimensions. The train-test split is 50}
    \label{fig:dimention}
\end{figure}

\textbf{Effect of dimension}.  We conduct investigations on the effect of dimension on the multi-Label classification task. For this goal, we change the embedding size of last ReLU layer in Fig \ref{fig:multimodal}. Figure \ref{fig:dimention} illustrates the results of Micro-F1 and Macro-F1 for BERTERS(ExEm(fastText)) by varying the number of dimensions. As the number of dimensions increase, the capable of storing more information becomes higher. Hence, We observe that the Micro and Macro values enhance as the number of dimensions rise.

\subsection{Visualization}
Figure \ref{visualization} shows the obtained results for the visualization task. Three different topics are used to color the nodes. Figures (a) to (c) respectively cluster experts based on ExEm(fastText), BERT, BERTERS(ExEm(fastText)) that is the concatenation of ExEm(fastText), BERT and normalized h-index and has 641 dimensions.  Although ExEm embeds experts farthest apart, but embeddings generated by BERTERS(ExEm(fastText)) well separate the communities. The reason is that three topics have overlaps, and a candidate can be expert in all of them. Thus the partition originated by this approach is more meaningful. In contrast, BERT embeds communities very closely. 

\textbf{Effect of dimension}.
Figures (c) to (f) illustrate the effect of dimension on visualization.  We make the observation that the performance of clustering improves as the number of dimensions grow. BERTERS(ExEm(fastText)) with dimensions 64 and 256 attempt to cluster experts with high intra-cluster edges together.  By comparison, BERTERS(ExEm(fastText)) with size 512 preserves the community structure better than low dimensions. Finally, the embeddings created from the concatenation can find the overlapping communities in which experts are interested in the same topic.

\subsection{Discussion}
The presented results in tables \ref{tb:macrof1} and \ref{tb:microf1} shows the comparison of our proposed multimodal approach using different setups. From these tables, BERTERS(ExEm(fastText)) and BERTERS(ExEm(Word2Vec)) are superior to other in terms of metrics. The reason behind this superiority is because of utilization ExEm. This method provides better graph presentation compared to others by using dominating set theories and thus, it is able to provide better results in various tasks such as classification. On the other hand, combination of document and graph modalities provides more accurate results because of adding extra textual information to existing method. According to our experiments, better presentation in both sides, text and graph, yields in better results for classification task but using which algorithm for presenting is hard choice and requires experiments to be evaluated.

\begin{figure}[ht]
\centering
    \subfloat[ExEm (dimension of embedding is 128)]{{\includegraphics[width=0.32\linewidth]{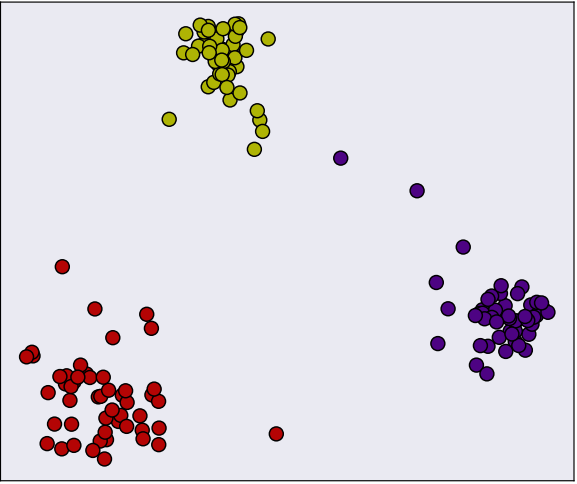} }}%
    \hfill
    \subfloat[BERT (dimension of embedding is 512)]{{\includegraphics[width=0.32\linewidth]{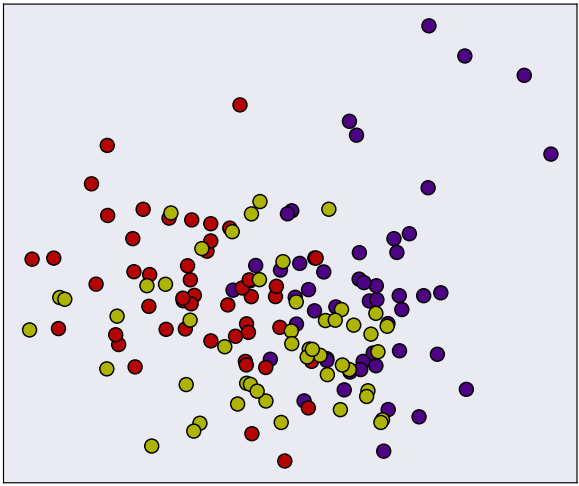} }}%
    \hfill
    \subfloat[BERTERS(ExEm(fastText)) (dimension of embedding is 641)]{{\includegraphics[width=0.32\linewidth]{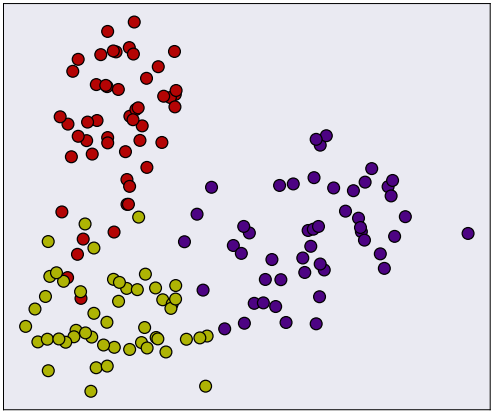} }}%
    \\
    \subfloat[BERTERS(ExEm(fastText)) (dimension of embedding is 64)]{{\includegraphics[width=0.32\linewidth]{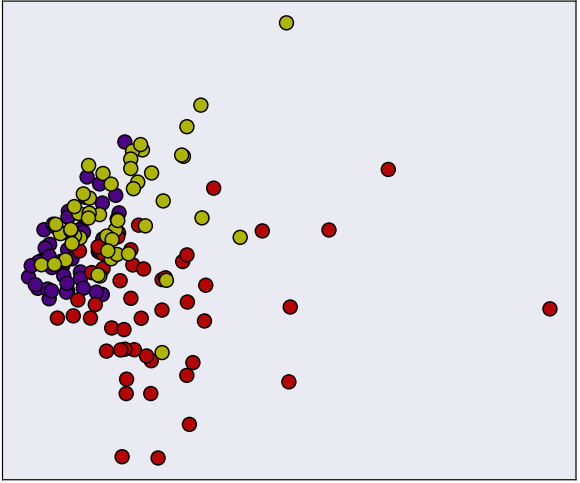}}}%
    \hfill
    \subfloat[BERTERS(ExEm(fastText)) (dimension of embedding is 256)]{{\includegraphics[width=0.32\linewidth]{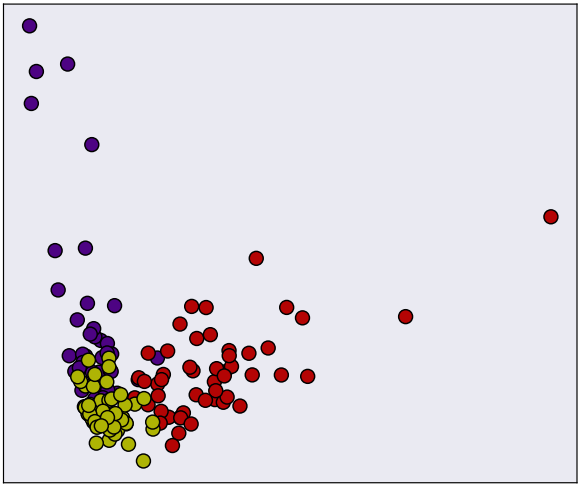} }}%
    \hfill
    \subfloat[BERTERS(ExEm(fastText)) (dimension of embedding is 512)]{{\includegraphics[width=0.32\linewidth]{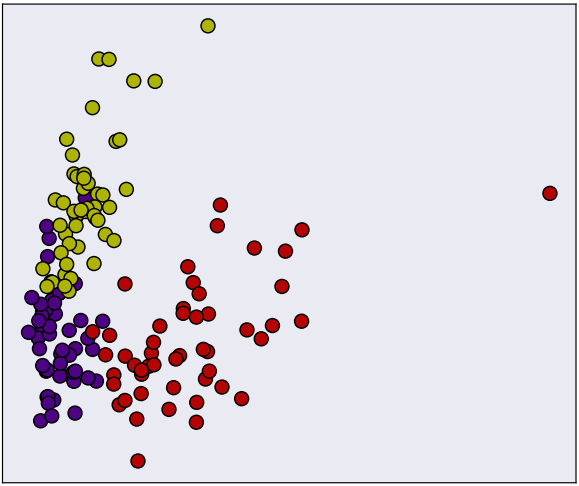} }}%
     \\
     \subfloat{{\includegraphics[width=0.3\linewidth]{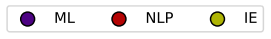} }}%
    \caption{Visualization of communities of 50 top experts in three topics for different techniques and dimensions. Each point corresponds to an expert. Color of an expert denotes its cluster.
}%
    \label{visualization}%
\end{figure}

Figure \ref{visualization} shows the visualization results for different setup that clearly from this figure is seen the embedding size is also another important hyper-parameter that affects the presentation. Part (a) from this figure shows using ExEm without document data that yields to completely separating three subject areas of ML, NLP and IE that we know is not correct due to the fact that there are significant overlaps among topics.  Having this in mind, and what is clearly seen from part (f), this separation is not done completely for authors who have been working in multiple fields such as IE and NLP together or any other combination of these three subject areas. A clear separation in this presentation is not always acceptable and for some hard cases such as what is shown in this figure, the inseparable subject areas must have collisions in some cases.

It is worth noting that gain in performance with increasing dimensions can be observed in both tasks.  As a conclusion, the best embeddings for finding experts in a ERS are directly generated from the concatenation of their values of normalized h-index, their presentations obtained from a co-author network by ExEm and experts' published items that converted into vectors by BERT. As mentioned in previous sections, it is possible to extend BERTERS into CQA to find the best users for answering the posted questions.

\begin{table}[ht]
\centering
\begin{tabular}{p{4.3cm}|p{0.96cm}|p{0.96cm}|p{0.96cm}|p{0.96cm}|p{0.96cm}|p{0.96cm}|p{0.96cm}|p{0.96cm}|p{0.96cm}}
\hline
Model & \multicolumn{9}{c}{Train ratio}\\
\cline{2-10}
& 0.1 & 0.2 & 0.3 & 0.4 & 0.5 & 0.6 & 0.7 & 0.8 & 0.9 \\
\hline
BERT & 0.6258 & 0.6565 & 0.6562 & 0.6624 & 0.6699 & 0.6602 & 0.6706 & 0.6718 & 0.6605\\
\hline
ExEm(fastText) & 0.5207 & 0.5309 & 0.5438 & 0.55 & 0.5628 & 0.5668 & 0.5655 & 0.5753 & 0.5769\\
\hline
ExEm(Word2Vec) & 0.5187 & 0.521 & 0.5489 & 0.5491 & 0.5604 & 0.5655 & 0.5631 &
0.5683 & 0.5686\\
\hline
BERTERS(DeepWalk) & 0.6497 & 0.6902 & 0.698 & 0.6973 & 0.6961 & 0.709 & 0.7098 & 0.7065 & 0.7129\\
\hline
BERTERS(Node2Vec) & \textbf{0.6648} & 0.6922 & 0.6951 & 0.7042 & 0.7042 & \textbf{0.714} & 0.7118 & 0.7025 & 0.7088 \\
\hline
BERTERS(ExEm(Word2Vec) & 0.6552 & 0.6906 & 0.6809 & \textbf{0.7047} & 0.7048 & 0.7042 & 0.7072 & 0.7058 & 0.7127 \\
\hline
BERTERS(ExEm(fastText)) & 0.6609 & \textbf{0.6924} & \textbf{0.6977} & 0.7001 & \textbf{0.7059} & 0.7129 & \textbf{0.7124} & \textbf{0.7141} & \textbf{0.7182} \\
\hline
\end{tabular}
\caption{Micro-F1 of multi-label classification task varying the train-test split ratio}\label{tb:microf1}
\end{table}

\begin{table}[ht]
\centering
\caption{Macro-F1 of multi-label classification task varying the train-test split ratio}\label{tb:macrof1}
\begin{tabular}{p{4.3cm}|p{0.96cm}|p{0.96cm}|p{0.96cm}|p{0.96cm}|p{0.96cm}|p{0.96cm}|p{0.96cm}|p{0.96cm}|p{0.96cm}}
\hline
Model & \multicolumn{9}{c}{Train ratio}\\
\cline{2-10}
& 0.1 & 0.2 & 0.3 & 0.4 & 0.5 & 0.6 & 0.7 & 0.8 & 0.9 \\
\hline
BERT & 0.4721 & 0.5201 & 0.5242 & 0.5306 & 0.5511 & 0.5284 & 0.533 & 0.5212 & 0.5361\\
\hline
ExEm(fastText) & 0.3728 & 0.3953 & 0.4031 & 0.4035 & 0.4231 & 0.4294 & 0.4357 & 0.4284 & 0.4269 \\
\hline
ExEm(Word2Vec) &0.3703 & 0.3783 & 0.4041 & 0.4062 & 0.4236 & 0.4253 & 0.4255 & 0.424 & 0.4227\\
\hline
BERTERS(DeepWalk) &0.5086 & 0.5686 & 0.5818 & 0.5747 & 0.5774 & 0.5876 & 0.5856 & 0.5785 & 0.5809\\
\hline
BERTERS(Node2Vec) & \textbf{0.5263} & \textbf{0.5707} & 0.5731 & \textbf{0.5841} & 0.58 & \textbf{0.5981} & 0.5768 & 0.587 & 0.5769\\
\hline
BERTERS(ExEm(Word2Vec) & 0.5137 & 0.5636 & 0.5712 & 0.5735 & 0.5803 & 0.5813 & 0.5819 & 0.5811 & 0.5878 \\
\hline

BERTERS(ExEm(fastText))& 0.5257 & 0.5702 & \textbf{0.5737} & 0.5743 & \textbf{0.5817} & 0.5836 & \textbf{0.5857} & \textbf{0.5883} & \textbf{0.5942} \\
\hline
\end{tabular}
\end{table}

\section{Conclusion} \label{sec:conclusion}
In this paper,  a multimodal classification approach, called BERTERS, has been proposed for expert recommendation system. In BERTERS, each candidate expert is represented by a vector which is the concatenation of three important features. One feature is the average of embeddings of all articles concerned with an expert. Each article converts to a vector by using BERT transformer. The second feature comes from applying a graph embedding technique on the co-author graph.  BERTERS uses three different graph embedding approaches including DeepWalk, Node2vec and ExEm. Finally,  normalized value of h-index is considered as third feature. Then, the concatenation of features is fed into a classifier that composes of three dense layers with ReLU function. In the final step, the performance of BERTERS was evaluated on the multi-label classification and visualization tasks and seven variants of the model. The results show that BERTERS(ExEm(fastText)) performs better than the other variants.

\bibliography{mybibfile}

\end{document}